\documentclass[runningheads]{llncs}
\usepackage[T1]{fontenc}
\usepackage[utf8]{inputenc}
\usepackage{microtype}
\usepackage{amsmath,amssymb}
\usepackage{graphicx,booktabs,array,tabularx}
\usepackage{xcolor}
\usepackage{changepage}
\usepackage{algorithm,algpseudocode}
\usepackage{multirow}
\usepackage[section]{placeins}
\usepackage{cite}
\usepackage{pifont,xspace}
\usepackage{pifont}
\newcommand{\mypara}[1]{\smallskip\noindent{\bf {#1}.} \xspace}
\usepackage{makecell}
\usepackage{bm}
\newcommand{\td}{$\mathsf{Train to Defend}~$}
\newcommand{\nospacetd}{$\mathsf{Train to Defend}$}
\newcommand{\ours}{{$\mathsf{SimBreak}$}\xspace}
\newcommand{\relu}{\operatorname{ReLU}}

\definecolor{deepred}{RGB}{200, 0, 0}
\newcommand{\no}{\textcolor{deepred}{\ding{55}}\xspace}
\definecolor{deepgreen}{RGB}{0, 137, 0}
\newcommand{\yes}{\textcolor{deepgreen}{\ding{51}}\xspace}
\usepackage[colorlinks=true,allcolors=blue]{hyperref}
\newcommand{\blackemail}[1]{%
\href{mailto:#1}{\textcolor{black}{#1}}%
}
\usepackage[nameinlink,noabbrev]{cleveref}

\usepackage{xcolor}
\begin{document}
\title{Breaking $\mathsf{TraintoDefend}$ via Adaptive Neuron Clustering in Cryptanalytic Model Extraction}
\titlerunning{ }
\author{Haojie Wu\inst{1} 
\and Haolin Liu\inst{2}
\and Tianshuo Cong\inst{1}
\and Guoyan Zhang\inst{1}}
\authorrunning{ }
\titlerunning{ }
\author{Haojie Wu\inst{1} 
\and Haolin Liu\inst{2}
\and Tianshuo Cong\inst{3}
\and Guoyan Zhang\inst{4}}
\authorrunning{ }
\institute{
School of Cryptologic Science and Engineering, Shandong University\\
\email{
\blackemail{202337050@mail.sdu.edu.cn}
}
\and
School of Computer Science, Shanghai Jiao Tong University\\
\email{\blackemail{liuhaolinsjtu@sjtu.edu.cn}}
\and
School of Cryptologic Science and Engineering, Shandong University\\
\email{
\blackemail{tianshuo.cong@sdu.edu.cn}
}
\and
School of Cryptologic Science and Engineering, Shandong University\\
\email{
\blackemail{guoyanzhang@sdu.edu.cn}
}
}   

\maketitle
\begin{abstract}
Cryptanalytic model extraction poses a fundamental threat to deep learning model confidentiality by enabling precise parameter recovery with theoretical guarantees of functional equivalence.
To counter this growing threat, \td (NeurIPS 2025) was proposed as the first dedicated defense against cryptanalytic extraction and remains the only defense believed to be effective to date.
It encourages neurons to learn similar weights so that they switch activation states together, making their individual contributions harder to distinguish.
However, we notice that neurons with similar weights need not switch activation states together under small perturbations.
Instead, weight similarity can cause the original clustering rule to merge signatures from distinct neurons when only a subset of their normalized weight coordinates agree.
Motivated by this observation, we introduce \ours, an adaptive clustering rule that breaks the weight-similarity defense of \nospacetd.
\ours estimates a relative scale from all jointly observed weight coordinates and uses the same scale to test both weight direction and bias consistency.
Across six protected fully connected networks trained on MNIST, \ours raises target layer signature recovery from $0\%$ to $100\%$, with a maximum relative recovery error of $2^{-22.27}$.
Our work challenges the use of neuron weight similarity as a sufficient basis for protection and establishes adaptive evaluation as essential to assessing cryptanalytic extraction defenses.

\end{abstract}

\section{Introduction}
\label{sec:introduction}
Deep Neural Networks (DNNs) constitute valuable digital assets, embodying considerable investment in proprietary data, computational resources, and engineering know-how.
To commercialize DNNs while preserving control over their parameters, providers typically offer them under the Model-as-a-Service (MaaS) paradigm, in which clients interact with remotely hosted models solely through inference APIs~\cite{maas2,maas3}. 
Nevertheless, recent progress in cryptanalytic model extraction~\cite{carlini_crypto_cea1,canales_eurocrypt_cea2,chen_asiacrypt_cea3} represents a severe threat.
Conventional learning-based extraction mainly approximates task behavior from sampled queries.
Cryptanalytic model extraction, by contrast, draws on cryptanalysis principles to design informative queries, thereby recovering hidden parameters with high precision and reconstructing functionally equivalent models, turning black-box access into model possession in effect.

Since Carlini et al.~\cite{carlini_crypto_cea1} demonstrated how the piecewise affine structure of ReLU networks can be exploited to recover model parameters through output queries, cryptanalytic model extraction has advanced along several directions.
Improvements in signature estimation and sign recovery have made extraction more efficient and enabled attacks on deeper networks~\cite{liu_eurocrypt_cea6}.
Meanwhile, studies of different network architectures~\cite{cryptoeprint:2026/1678-transformer,sun_acisp_cea7,li_eprint_cea8,cryptoeprint:2026/464-max-pooling-liu,cryptoeprint:2026/1164-cnn-maxpooling-chen} and activation functions~\cite{cryptoeprint:2026/178-nonlinear,cryptoeprint:2026/253-nonlinear} have explored how structural properties and nonlinear behavior reveal hidden parameters.
Moreover, attacks can recover parameter information from hard labels~\cite{chen_asiacrypt_cea3,carlini_eurocrypt_cea4,cryptoeprint:2026/2061-Normal-Alignment-Improved-Cryptanalytic-Sign-Recovery-on-Hard-Label-Networks,cryptoeprint:2026/2097-Efficient-Sign-Recovery} alone by examining decision boundaries.
These advances broaden the range of models and interfaces exposed to cryptanalytic extraction, motivating defenses that account for increasingly capable adversaries.

To counter this threat, $\mathsf{TraintoDefend}$\cite{kurian_neurips_ced1} was introduced as the first dedicated defense against cryptanalytic extraction, making its robustness a central question for the protection of model parameters.
\td modifies the training objective to encourage different neurons to learn similar weights.
The defense rests on the premise that similar weights cause neurons to switch activation states together, mixing their contributions to local derivative measurements and hindering signature recovery.
Because existing attacks recover layers sequentially, the defense can be applied to the first layer alone.
The authors report that existing attacks fail to extract the evaluated models within a $48$-hour budget.
\Cref{bg:t2d} describes the defense in detail.

However, we highlight that extraction involves several stages, from locating critical points and estimating local signatures to grouping observations and reconstructing individual neurons.
The failure of the overall procedure does not by itself establish which stage is affected or whether the obstacle persists under a modified attack.
If the defense leaves recoverable signatures accessible but causes the existing implementation to group them incorrectly, its reported protection can be defeated by changing how the attacker processes those observations.
Determining whether this is the case directly tests the claimed barrier to parameter recovery.
This raises the following question:
\begin{adjustwidth}{0.5cm}{0.5cm}
\begin{center}
\textit{Does weight similarity really guarantee security against cryptanalytic model extraction?}
\end{center}
\end{adjustwidth}

\subsection{Our Work}
\label{sec:contributions}

\mypara{Theoretical analysis of weight similarity in cryptanalytic extraction}
We characterize when a single perturbation causes multiple neurons to switch activation states in~\Cref{sec:activation-changes}.
Joint switching depends on the probe center, direction, and step size, so weight similarity alone does not ensure that a probe crosses multiple activation boundaries.
We identify signature clustering as a concrete obstacle introduced by \td in the existing extraction attack.
As analyzed in~\Cref{sec:clustering-failure}, the original rule can merge signatures from distinct neurons when sufficiently many normalized weight coordinates agree, even if the remaining coordinates differ substantially.
Graph connectivity then propagates these incorrect matches into mixed clusters, hindering the recovery of individual neuron signatures.

\mypara{\textbf{\textsf{SimBreak}} for distinguishing neurons with similar weights}
To overcome the clustering obstacle introduced by \nospacetd, \ours{} formulates pairwise signature matching as a test of whether two local observations agree under a common relative scale.
It estimates this scale by least squares over all jointly observed weight coordinates, as described in~\Cref{sec:observation-alignment}.
The resulting normalized residual incorporates discrepancies across every jointly observed coordinate, preventing agreement on a small subset from being sufficient for a match.
The rule also applies the fitted scale to the bias estimates, providing an additional check for signatures with nearly collinear weight vectors.

\mypara{Experimental evaluation of signature recovery with \textbf{\textsf{SimBreak}}}
We evaluate \ours{} on six models with varying widths and depths and compare it with the original coordinate agreement rule.
Replacing the original rule with \ours{} raises first-layer signature recovery from $0\%$ to $100\%$ across all six models, with a maximum reported relative recovery error of $2^{-22.27}$.
Complete recovery is achieved across widths from $8$ to $64$ in two-hidden-layer networks and also in the deeper models, as shown in~\Cref{sec:experiments:rq3}.
These results demonstrate that \ours{} overcomes the clustering barrier imposed by \td and restores recovery of the signatures that the defense seeks to protect.

In summary, our main contributions are as follows.
\begin{itemize}
\item A theoretical analysis exposes a weakness in \nospacetd's protection mechanism and identifies signature misclustering as a concrete barrier to the existing attack.

\item To overcome this barrier, we propose \ours{}, an adaptive clustering rule that jointly evaluates weight direction and bias consistency under a relative scale fitted over all commonly observed weight coordinates.
\item \ours{} bypasses the protected layer signature recovery barrier of \td, raising recovery from $0\%$ to $100\%$ across all six evaluated models with a maximum reported relative error of $2^{-22.27}$.
\end{itemize}

\subsection{Organization.}
\label{sec:organization}
The paper proceeds from the threat model and extraction background in \Cref{sec:background}, through the failure analysis in \Cref{sec:failure}, to the adaptive rule in \Cref{sec:method}.
\Cref{sec:experiments} presents the experimental evaluation.
\Cref{sec:discussion} discusses what these findings imply for defense evaluation.

% \subsubsection{Code Available.}The source code of this work will be available at \url{xxx}.

\section{Background}
In this section, we introduce the basic notation and definitions used throughout this work, largely following the conventions adopted in prior works~\cite{carlini_crypto_cea1,liu_eurocrypt_cea6}.
\label{sec:background}
\subsection{Basic Notation and Definitions}

\begin{definition}[ReLU-Fully Connected Network]
\label{def:network}
A fully connected network with $L$ hidden layers is a function $f_\theta:\mathbb R^{d_0}\to\mathbb R^{d_{L+1}}$ with parameters $\theta=\{W^{(\ell)},b^{(\ell)}\}_{\ell=1}^{L+1}$ and composition
\begin{equation*}
f_\theta=f_{L+1}\circ\sigma\circ f_L\circ\cdots\circ\sigma\circ f_1.
\end{equation*}
Given a network input $x$, the affine map in layer $\ell$ yields $z^{(\ell)}(x)=f_\ell(h^{(\ell-1)}(x))=W^{(\ell)}h^{(\ell-1)}(x)+b^{(\ell)}$, where $h^{(0)}(x)=x$, $W^{(\ell)}\in\mathbb R^{d_\ell\times d_{\ell-1}}$, and $b^{(\ell)}\in\mathbb R^{d_\ell}$.
The activation function $\sigma(x)$ acts on each component after each hidden layer; the output layer is affine.
\end{definition}
In this work, we consider the $\relu$ as the activation function, i.e., $\sigma(t)=\relu(t)=\max\{0,t\}$, which has been the most extensively studied among these lines of research.

\begin{definition}[Hidden Layer and Neuron]
\label{def:neuron}
Starting from $h^{(0)}(x)=x$, the pre-activation and activation of hidden layer $\ell$ are
\begin{equation*}
z^{(\ell)}(x)=W^{(\ell)}h^{(\ell-1)}(x)+b^{(\ell)},
\quad h^{(\ell)}(x)=\sigma\bigl(z^{(\ell)}(x)\bigr).
\end{equation*}
Neuron $j$ has pre-activation $z_j^{(\ell)}(x)=W_{j,:}^{(\ell)}h^{(\ell-1)}(x)+b_j^{(\ell)}$ and activation $h_j^{(\ell)}(x)=\sigma(z_j^{(\ell)}(x))$.
It is active when $z_j^{(\ell)}(x)>0$ and inactive otherwise.
\end{definition}

\begin{definition}[Critical Point and Critical Set]
\label{def:critical}
An input $x^*\in\mathcal X$ is a critical point of neuron $j$ in layer $\ell$ if $z_j^{(\ell)}(x^*)=0$.
Its critical set on the input domain $\mathcal{X}$ is $\mathcal C_j^{(\ell)}=\{x\in\mathcal X:z_j^{(\ell)}(x)=0\}$.
\end{definition}

\begin{definition}[Signature]
\label{def:signature}
A signature of neuron $j$ in hidden layer $\ell$ consists of a weight and bias with a shared scale:
\begin{equation*}
\widehat W_{j,:}^{(\ell)}=\alpha W_{j,:}^{(\ell)},\quad\widehat b_j^{(\ell)}=\alpha b_j^{(\ell)}.
\end{equation*}
\end{definition}

\subsection{Cryptanalytic Extraction Attacks}

Cryptanalytic extraction reconstructs fully connected ReLU networks by exploiting their piecewise affine structure through queries that return raw numerical outputs or hard labels~\cite{carlini_crypto_cea1,chen_asiacrypt_cea3,carlini_eurocrypt_cea4}.
The attack proceeds layer by layer, recovering neuron signatures and then determining their signs before moving to the next layer.
Since \td modifies the similarity of weights, we focus on the observation and clustering steps of signature recovery in this section.

\mypara{Step-I: Locating Critical Points}
The attacker first searches for critical points where a neuron changes its activation state.
Specifically, the attacker probes the network along a chosen input line, on which the output is piecewise affine.
Differences between the slopes estimated on nearby intervals indicate a transition between affine regions.
The attacker then narrows the interval containing the transition through repeated subdivision and additional queries.
This refinement localizes a critical point, around which finite difference measurements can be used to estimate the associated neuron's parameters~\cite{carlini_crypto_cea1,foerster_neurips_cea5}.

\mypara{Step-II: Signature Recovery}
At an isolated critical point, finite difference measurements on either side reveal the change in the local derivative.
For a neuron in the first layer, this change is proportional to its weight.
For deeper neurons, the attacker uses the recovered preceding layers to interpret the measurements in the coordinates of the target layer.

These measurements determine weight ratios rather than absolute parameter values.
The unknown scale reflects the contribution of the subsequent network to the observed output change.
Signature recovery combines this information to reconstruct the weights and associated bias up to scale.
Positive homogeneity of ReLU allows this scaling to be compensated by dividing the weight $W_{:,j}^{(\ell+1)}$ by $\alpha$~\cite{carlini_crypto_cea1}.
Local derivative measurements initially determine the scale only up to an unknown sign, which is resolved during sign recovery.
The information obtained at each critical point is represented as a local signature observation.

\begin{definition}[Signature Observation]
\label{def:observation}
A local observation at a critical point $x_p$ of a neuron in hidden layer $\ell$ consists of a weight estimate $\widehat W^{[p]}$, an associated bias estimate $\widehat b^{[p]}$, and observed coordinates $\Omega_p\subseteq\{1,\ldots,d_{\ell-1}\}$.
For an observation from neuron $i$, these estimates satisfy
\begin{equation*}
\widehat W^{[p]}_{\Omega_p}\approx s_pW_{i,\Omega_p}^{(\ell)},\quad\widehat b^{[p]}\approx s_pb_i^{(\ell)},\quad s_p\ne0.
\end{equation*}
The same scale $s_p$ applies to the weight and bias and may have either sign.
For any two observations $p,q$ from the same layer, their common observed coordinates are $\Omega_{p,q}=\Omega_p\cap\Omega_q$.
\end{definition}
In deeper layers, if some neurons in the preceding layer remain inactive near the critical point, their outputs are locally zero, so the corresponding weights cannot be inferred from that local signature observation.
These unresolved coordinates are treated as temporarily missing.

The attacker therefore collects observations at multiple critical points.
Points belonging to the same neuron may occur under different activation patterns of the preceding layers, exposing complementary subsets of its weights.
Because the neuron identities and scales of these observations are unknown, the attacker clusters them before combining their recovered coordinates.

\begin{definition}[Signature Clustering based on Coordinate Agreement]
\label{def:clustering}
For observations $p,q$ from the same layer, choose a common coordinate $a\in\Omega_{p,q}$ with nonzero estimated weights and normalize to eliminate the unknown scaling factor $s_p,s_q$:
\begin{equation*}u_k^{[p,a]}=\frac{\widehat W_k^{[p]}}{\widehat W_a^{[p]}},\quad u_k^{[q,a]} =\frac{\widehat W_k^{[q]}}{\widehat W_a^{[q]}}.
\end{equation*}
The relative discrepancy at coordinate $k$ is
\begin{equation*}
e_k^{[p,q,a]}=\frac{|u_k^{[p,a]}-u_k^{[q,a]}|}{|u_k^{[p,a]}|}+\frac{|u_k^{[p,a]}-u_k^{[q,a]}|}{|u_k^{[q,a]}|}.
\end{equation*}
Let $\Omega_{p,q}^{(a)}$ contain the common coordinates for which this comparison is finite and well defined.
The number of agreeing coordinates is
\begin{equation*}
M_a(p,q)=\sum_{k\in\Omega_{p,q}^{(a)}}\mathbf 1\!\left\{e_k^{[p,q,a]}<\epsilon_{\mathrm{coord}}\right\}.
\end{equation*}
The rule connects observations $p,q$ when $M_a(p,q)\geq\tau_{\mathrm{count}}$ for at least one tested normalization coordinate $a$.
The connected components form candidate neuron weight clusters.
\end{definition}

Within each cluster, the attacker estimates the relative scales of the observations from their overlapping coordinates.
The aligned observations are then combined to recover coordinates that were missing at individual critical points.
Multiple estimates of the same coordinate also provide redundancy for reducing numerical error~\cite{carlini_crypto_cea1,foerster_neurips_cea5}.
If a candidate signature remains incomplete, the attacker collects further critical points and updates the grouping.
The recovered direction and a corresponding critical point determine the bias under the same scale.

\mypara{Step-III: Sign Recovery}
It leaves an orientation ambiguity in the Signature Recovery step: multiplying both the weight and bias by a negative scalar preserves their ratios but reverses the active side of the ReLU.
Sign recovery aims to resolve this ambiguity.
Early extraction attacks used an expensive search over sign assignments, while subsequent work developed more effective procedures for recovering these signs\cite{carlini_crypto_cea1,canales_eurocrypt_cea2,foerster_neurips_cea5,cryptoeprint:2026/2061-Normal-Alignment-Improved-Cryptanalytic-Sign-Recovery-on-Hard-Label-Networks,cryptoeprint:2026/2097-Efficient-Sign-Recovery}.

\mypara{Step-IV: Layer-by-Layer Peeling}
Once the signatures and signs of the target layer have been recovered, the attacker can compute its activations for each queried input.
These activations serve as known inputs to the next layer, allowing the attacker to repeat the above operations for the next layer~\cite{carlini_crypto_cea1,liu_eurocrypt_cea6}.
The process continues through the hidden layers, after which the final affine layer is recovered by solving a linear system using the recovered hidden activations and oracle outputs.

\subsection{\td}\label{bg:t2d}

Kurian and Aysu~\cite{kurian_neurips_ced1} proposed \td as the first dedicated defense against cryptanalytic parameter extraction.
The defense targets a premise of signature recovery: when a local perturbation crosses an activation boundary associated with a single neuron, the resulting change in the model's output derivatives can reveal that neuron's weight direction.
If the same probe crosses several activation boundaries, the measured change can instead combine contributions from multiple neurons, making it harder to attribute the measurement to any one of them.
This motivates the design of \nospacetd, which encourages neurons to learn similar weights with the aim of making them switch activation states together.

For a scalar model output, consider a perturbation in the input space of layer $\ell$, with $u=h^{(\ell-1)}(x)$.
Suppose that the activation patterns in subsequent layers remain fixed and that only neurons $i$ and $j$ change states across the probe.
The resulting difference between the output gradients with respect to $u$ is
\begin{equation*}
\Delta\nabla_u f_\theta = s_i c_i W_{i,:}^{(\ell)} + s_j c_j W_{j,:}^{(\ell)},
\end{equation*}
where $c_i$ and $c_j$ are the effective downstream coefficients and $s_i,s_j\in\{-1,1\}$ indicate the directions of the state changes.
When only neuron $i$ changes state and its contribution is nonzero, the gradient difference reveals its weight direction up to sign.
When both neurons change states, the gradient difference combines their contributions and can obscure their individual weight directions.
Motivated by this observation, \td augments the task loss with a regularizer that encourages selected neuron pairs to learn similar weights.

Specifically, \td augments the task loss with a regularization term that penalizes weight differences between selected neuron pairs:
\begin{equation}\label{eq-t2d}
\mathcal L=\mathcal L_{\mathrm{task}}+\lambda_{\mathrm{sim}}\sum_{(i,j)\in\mathcal P}\left\|W_{i,:}^{(\ell)}-W_{j,:}^{(\ell)}\right\|_2^2,
\end{equation}
where $\mathcal P$ contains the selected neuron pairs in the protected layer $\ell$, and $\lambda_{\mathrm{sim}}$ controls the strength of the regularizer.
The two terms serve different objectives: the task loss rewards predictive performance, while the similarity penalty reduces the distance between the weight vectors of the selected pairs.
The defense controls this trade-off through both the regularization strength and the fraction of pairs included in $\mathcal P$.
The regularizer acts on the weights during training, so the defense requires no additional operation at inference time.
The authors also consider applying the penalty only to the first layer, corresponding to $\ell=1$ in~\Cref{eq-t2d}.
This choice exploits the sequential structure of existing extraction attacks, which use the recovered parameters of earlier layers to compute the inputs needed for recovering later ones.

Weight similarity can affect two distinct stages of the attack.
During local signature estimation, the relevant question is whether a single probe actually crosses multiple activation boundaries.
During clustering, the attack must decide whether signatures estimated at different critical points originate from the same neuron.
The original extraction attack matches two signature observations when sufficiently many of their normalized weight coordinates agree within a prescribed tolerance.
Agreement on sufficiently many coordinates can therefore cause the rule to merge observations from distinct neurons, even when their activation boundaries can be probed separately.
\Cref{sec:failure} analyzes both mechanisms and examines how they arise.

\subsection{Other Defenses}
Noman et al~\cite{mit65610_2026_rounding}. investigate output rounding as a defense against cryptanalytic extraction, reducing the precision of returned outputs while leaving the model parameters unchanged.
Rounding suppresses the small output differences used by the original attack, disrupting critical point search and local derivative estimation.
However, their adaptive attack recovers models by estimating directional derivatives over larger probing intervals.
Moreover, cryptanalytic extraction has also been demonstrated in the hard-label setting, where the attacker observes only predicted class labels~\cite{chen_asiacrypt_cea3,carlini_eurocrypt_cea4}.
Together, these results show that restricting output precision alone does not generally prevent cryptanalytic extraction.
Against this backdrop, \td emerged as the first, and thus far the only, defense believed to be effective against cryptanalytic model extraction.

\subsection{Threat Model}
\label{sec:goals}
We follow the threat model considered by \nospacetd, and prior cryptanalytic extraction attacks~\cite{kurian_neurips_ced1,carlini_crypto_cea1,foerster_neurips_cea5}.
A model provider trains and deploys a fully connected ReLU network $f_\theta$, and an adversary attempts to recover its parameters through oracle queries.
We consider an adaptive adversary that knows the defense mechanism and adjusts the extraction procedure to bypass its protection.

\mypara{Defender Goals and Resources}
The defender controls the training process and may modify the training objective to resist cryptanalytic parameter extraction.
It applies \td to encourage similarity between the weights of selected neurons while maintaining predictive performance.
The defender chooses the regularization strength, the protected layers, and the neuron pairs subject to regularization.
Its goal is to prevent the adversary from recovering the model parameters through inference queries.
The protection is embedded in the trained parameters, while the network architecture and inference interface remain unchanged.
We consider protection embedded in the trained model, with no additional protective mechanism during inference.
Accordingly, query filtering, rate limiting, output perturbation, and access monitoring are outside this setting.
This allows us to examine whether the similar weights induced by \td can resist an adaptive extraction under the same query interface.

\mypara{Adversarial Goals and Resources}
The adversary has adaptive query access to an oracle $\mathcal O(x)=f_\theta(x)$.
It knows the network architecture and activation functions and may submit arbitrary inputs from $\mathcal X=\mathbb R^{d_0}$.
We consider raw numerical output access, under which the oracle returns the complete model output.
The adversary has no access to the target parameters, training data, or intermediate activations.

The adversary knows the mechanism of \nospacetd, and may exploit this knowledge when designing the attack.
In particular, it may adapt the comparison and clustering of local observations to account for the neuron weight similarity induced by the defense.
The adversary aims to recover parameters $\widehat\theta$ that yield a functionally equivalent model~\cite{carlini_crypto_cea1,jagielski_usenix_tea3}.
Parameter recovery accounts for neuron permutations and positive scaling transformations that preserve the network's function.
For an evaluation distribution $\mathcal D$ over $\mathcal X$, the extracted model is $(\epsilon,\delta)$-functionally equivalent to the target if
\begin{equation*}
\Pr_{x\sim\mathcal D}\left[\left\|f_\theta(x)-f_{\widehat\theta}(x)\right\|_2\leq\epsilon\right]\geq1-\delta.
\end{equation*}

\section{Revisiting \td}
\label{sec:failure}
\td motivates its defense by arguing that neurons with similar weights can switch activation states together, causing local derivative measurements to combine their contributions.
Our key insight is that weight similarity alone does not determine which neurons switch under a given probe, yet it can directly affect how the original attack groups the resulting signature observations.
We analyze these two effects separately.

\subsection{Conditions for Joint Activation Changes}
\label{sec:activation-changes}

Consider a critical point $x$ of neuron $i$ in layer $\ell$, so that $z_i^{(\ell)}(x)=0$.
Let $\delta$ be a perturbation direction and $\varepsilon>0$ the probe step size.
Suppose that the activation patterns in layers $1,\ldots,\ell-1$ remain fixed along the probe segment $\{x+t\delta:|t|\leq\varepsilon\}$.
The map from the network input to the input of layer $\ell$ is affine along this segment, giving
\begin{equation*}
h^{(\ell-1)}(x+t\delta)=h^{(\ell-1)}(x)+tA^{(\ell-1)}\delta,
\end{equation*}
where $A^{(\ell-1)}$ is the linear part of the affine map and $A^{(0)}=I$.
The preactivation of another neuron $j$ at the two endpoints is therefore
\begin{equation*}
z_j^{(\ell)}(x\pm\varepsilon\delta)=z_j^{(\ell)}(x)\pm\varepsilon W_{j,:}^{(\ell)}A^{(\ell-1)}\delta.
\end{equation*}
These endpoint preactivations have strictly opposite signs if and only if
\begin{equation*}
\left|z_j^{(\ell)}(x)\right|<\varepsilon\left|W_{j,:}^{(\ell)}A^{(\ell-1)}\delta\right|.
\end{equation*}
For a probe that crosses the critical boundary of neuron $i$, this inequality determines whether it also strictly crosses the boundary of neuron $j$.
The left-hand side measures how far the probe center is from zero preactivation for neuron $j$, while the right-hand side measures how much the probe can change that preactivation.
If $z_j^{(\ell)}(x)\ne0$, a sufficiently small step along the fixed direction avoids crossing neuron $j$'s boundary.
Thus, similarity between the weight rows of neurons $i$ and $j$ alone does not establish that the same probe crosses both boundaries.
We will evaluate how often probes around the critical points found by the attack cross multiple activation boundaries in~\Cref{sec:experiments:rq1}.

\subsection{Why Weight Similarity Can Cause Clustering Errors}
\label{sec:clustering-failure}
We next analyze how the signature observations are processed.
The next procedure clusters observations believed to originate from the same neuron and combines them to reconstruct its signature.
Our inspection identifies a failure mechanism in this association step: observations from distinct neurons can satisfy the original matching criterion and enter the same cluster.
\subsubsection{Clustering Based on Coordinate Agreement.}
The original attack groups local signature observations using the coordinate agreement rule in \Cref{def:clustering}.
For each tested normalization coordinate, the rule counts how many common coordinates agree within a relative tolerance.
For observations of dimension $d$, the used count threshold
\begin{equation*}
\tau_{\mathrm{count}}=\max\{4,2(\log d-2)\}.
\end{equation*}
At $d=784$, agreement on ten valid coordinates suffices, including the normalization coordinate when its comparison is valid.
Once the count threshold is met, discrepancies on the remaining coordinates do not prevent a match.

\subsubsection{Effect of Weight Similarity.}
Consider distinct neurons $i$ and $j$ in layer $\ell$ with $W_{j,:}^{(\ell)}=W_{i,:}^{(\ell)}+\Delta W_{j,i}$.
For a normalization coordinate $a$ at which both weights are nonzero, the difference between the normalized weights at coordinate $k$ is
\begin{equation*}
\frac{W_{j,k}^{(\ell)}}{W_{j,a}^{(\ell)}}-\frac{W_{i,k}^{(\ell)}}{W_{i,a}^{(\ell)}}=\frac{W_{i,a}^{(\ell)}\Delta W_{j,i,k}-W_{i,k}^{(\ell)}\Delta W_{j,i,a}}{W_{i,a}^{(\ell)}\bigl(W_{i,a}^{(\ell)}+\Delta W_{j,i,a}\bigr)}.
\end{equation*}
When the normalization weights and the compared weights are bounded away from zero, sufficiently small differences at coordinates $a$ and $k$ make the relative discrepancy at coordinate $k$ small.
If this occurs on enough coordinates, the rule can match observations from distinct neurons even when their full weight vectors are not proportional.
The attack then forms clusters as connected components of the matching graph.
A single erroneous edge can join two previously separate components, even when most pairs of observations across the two components would fail the matching test.
Weight similarity can therefore disrupt signature recovery through incorrect associations even when individual activation boundaries can be probed separately.
We will evaluate cross-neuron matches and the resulting mixed clusters in~\Cref{sec:experiments:rq2}.

\section{\textbf{\textsf{SimBreak}}: Clustering beyond Coordinate Agreement}
\label{sec:method}
Our analysis shows that coordinate agreement can cause the original clustering rule to associate observations from different neurons.
We introduce \ours: an adaptive criterion that checks whether a single relative scale consistently aligns the observations over their common coordinates, and also requires biases to agree under the same scale.
The resulting rule uses consistency across the observed weights and biases to distinguish neurons with similar weights.

\subsection{Aligning Local Observations}
\label{sec:observation-alignment}

Local observations recovered at different critical points may have different scales, even when they originate from the same neuron.
Before comparing two observations, we therefore estimate their relative scale over the coordinates available in both.
For observations $p$ and $q$, let
$\Omega_{p,q}=\Omega_p\cap\Omega_q.$
We require $|\Omega_{p,q}|\geq m_{\min}$ and exclude unobserved coordinates from the comparison.

If both observations originate from neuron $i$ in layer $\ell$, their weight estimates satisfy
\begin{equation*}
\widehat W_{\Omega_{p,q}}^{[p]}\approx s_p W_{i,\Omega_{p,q}}^{(\ell)},\quad\widehat W_{\Omega_{p,q}}^{[q]}\approx s_q W_{i,\Omega_{p,q}}^{(\ell)}.
\end{equation*}
Consequently, a single relative scale $\lambda=s_p/s_q$ should align the two estimates across their common coordinates.
We estimate this scale by minimizing the squared alignment residual:
\begin{equation*}
\lambda^*=\underset{\lambda\in\mathbb R}{\operatorname{argmin}}\left\|\widehat W_{\Omega_{p,q}}^{[p]}-\lambda\widehat W_{\Omega_{p,q}}^{[q]}\right\|_2^2.
\end{equation*}
When $\widehat W_{\Omega_{p,q}}^{[q]}$ has nonzero norm, the solution is
\begin{equation*}
 \lambda^*
 =
 \frac{
 \left\langle
 \widehat W_{\Omega_{p,q}}^{[p]},
 \widehat W_{\Omega_{p,q}}^{[q]}
 \right\rangle
 }{
 \left\|\widehat W_{\Omega_{p,q}}^{[q]}\right\|_2^2
 }.
\end{equation*}
The estimated scale may be positive or negative because the signs of the local observations have not yet been resolved.
This alignment uses all common coordinates rather than dividing by a single selected weight.
We next use $\lambda^*$ to assess direction consistency and apply the same scale to check bias consistency.

\subsection{Clustering based on Direction and Bias Matching}
\label{sec:direction-bias}

Using the estimated relative scale $\lambda^*$, we compare the observations through their weight directions and biases.
Observations from the same neuron should agree in both quantities under this shared scale.

\subsubsection{Direction Matching.}
For weight estimates on the common coordinates, we define the distance of direction as
\begin{equation*}
D_{\mathrm{dir}}(p,q)= \frac{\left\|\widehat W_{\Omega_{p,q}}^{[p]}-\lambda^*\widehat W_{\Omega_{p,q}}^{[q]}\right\|_2}{\left\|\widehat W_{\Omega_{p,q}}^{[p]}\right\|_2}.
\end{equation*}
This normalized distance measures how closely the two observations align across all common coordinates.
Unlike the original coordinate count, it incorporates discrepancies on every compared coordinate.
The observations pass the direction test when
$D_{\mathrm{dir}}(p,q)\leq\tau_{\mathrm{dir}}$.

Since $\lambda^*$ may have either sign, the test accepts both parallel and antiparallel observations.
In particular, if $\varphi$ is the angle between the restricted weight estimates, then
\begin{equation*}
 D_{\mathrm{dir}}(p,q)
 =
 \sqrt{1-\cos^2\varphi}.
\end{equation*}
The comparison therefore does not require resolving the signs before clustering.

\subsubsection{Bias Matching.}
Direction matching alone cannot distinguish neurons whose critical hyperplanes are parallel but have different offsets (i.e., different bias)\footnote{The attacker can directly reuse a recovered signature for all neurons with identical weights and biases to preserve the network's function.
If the grouping is ambiguous, the attacker can enumerate candidate groupings, recover the remaining layers for each, and compare the resulting networks with the target on the same inputs.
This search increases extraction cost but can yield a functionally equivalent network.}.
We therefore compare the bias distance using the same scale obtained from the weights:
\begin{equation*}
D_{\mathrm{bias}}(p,q)=\frac{\left|\widehat b^{[p]}-\lambda^*\widehat b^{[q]}\right|}{\left|\widehat b^{[p]}\right|+\left|\lambda^*\widehat b^{[q]}\right|}.
\end{equation*}
The observations pass the bias test when
$D_{\mathrm{bias}}(p,q)\leq\tau_{\mathrm{bias}}$.
We accept a pair (i.e., they should be clustered into a group) only when both tests pass:
\begin{equation*}
 D_{\mathrm{dir}}(p,q)\leq\tau_{\mathrm{dir}},
 \quad
 D_{\mathrm{bias}}(p,q)\leq\tau_{\mathrm{bias}}.
\end{equation*}

\section{Experiment}
\label{sec:experiments}
We evaluate the behavior of \td and the effectiveness of \ours through the following research questions (\textbf{RQs}):
\begin{itemize}
\item \textbf{RQ1:} Does \td cause multiple neurons to change activation states around the critical points?
\item \textbf{RQ2:} How does \td affect signature clustering in the original extraction attack?
\item \textbf{RQ3:} Can \ours recover the signatures of neurons in models protected by $\mathsf{TraintoDefend}$?
\end{itemize}
\subsection{Experimental Setup}
\label{sec:experimental-setup}

\subsubsection{Implementation and Models.}
We use the publicly released implementation of \td and the public models provided with it~\cite{kurian_neurips_ced1}.
The models are fully connected networks with ReLU hidden layers trained on MNIST~\cite{MNIST}, with $784$ input dimensions and a scalar output.
The activation experiments use protected models with hidden layers $8\times2$, $16\times2$, $16\times3$, $16\times8$, $32\times2$, and $64\times2$.

\subsubsection{Evaluation Procedure.}
We focus on the signature recovery of the first hidden layer, which is also the configuration recommended by $\mathsf{TraintoDefend}$.
For \textbf{RQ1}, we run the public critical point search and inspect the activation states around the returned points using the model parameters.
At each critical point, we use $1{,}571$ perturbation directions and evaluate the same directions at different finite difference step sizes.
We count the probes in which at least two neurons in the first layer change activation states.
For \textbf{RQ2}, we examine the clusters produced by the original attack and record the neuron identities involved in incorrect clustering and the erroneous connections.
For \textbf{RQ3}, we evaluate signature recovery after replacing the original matching rule with \ours{}.
We report the number of recovered neurons and the relative error of the recovered weights $\theta$ after accounting for scale and sign ambiguity: $\frac{\max |\theta'-\theta|}{|\theta|}$.

\subsection{RQ1: Activation Changes around Critical Points}\label{sec:experiments:rq1}

\begin{figure}[b]
\centering
\includegraphics[width=0.9\linewidth]{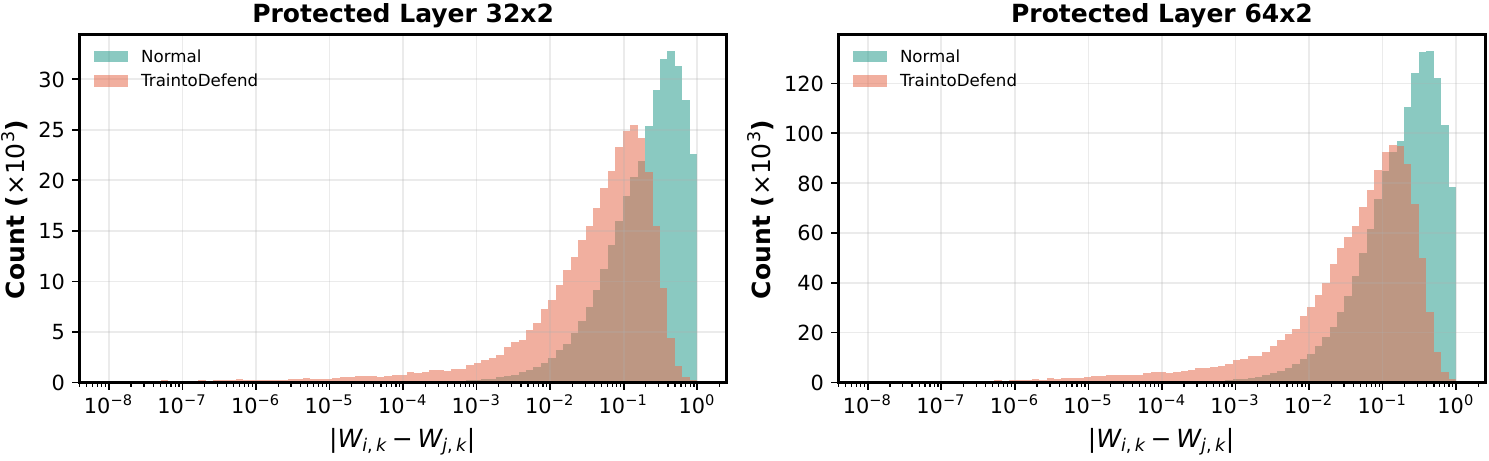}
\caption{Distributions of per-coordinate weight differences in the first layer of different models trained on MNIST.
Distributions for protected models extend further toward small differences.}
\label{fig:weightdifferences}
\end{figure}

\begin{table}[t]
\centering
\setlength{\tabcolsep}{3pt}
\caption{Number of probes crossing multiple first layer activation boundaries in models protected by $\mathsf{TraintoDefend}.$
$\bm N_{\mathrm{pts}}$: the number of critical points, $\bm N_{\mathrm{probe}}$: the total number of probes at each step size.
At each critical point, we use $1571$ perturbation directions.}
\label{tab:joint-activation}
\begin{tabular}{crrrrrrr}
\toprule
\multirow{2}{*}{\bfseries Model}
& \multirow{2}{*}{$\bm N_{\mathrm{pts}}$}
& \multirow{2}{*}{\bfseries$\bm N_{\mathrm{probe}}$}
& \multicolumn{5}{c}{\bfseries Step Size $\bm \varepsilon$} \\
\cmidrule(lr){4-8}
& & & $10^{-5.5}$ & $10^{-5}$ & $10^{-4.5}$ & $10^{-4}$ & $10^{-3}$ \\
\midrule
${784\text{-}8\times2\text{-}1}$  & $74$  & $116254$ & $0$ & $0$ & $0$ & $0$  & $715$ \\
${784\text{-}16\times2\text{-}1}$ & $187$ & $293777$ & $0$ & $0$ & $0$ & $0$  & $502$ \\
${784\text{-}16\times3\text{-}1}$ & $192$ & $301632$ & $0$ & $0$ & $0$ & $0$  & $19$ \\
${784\text{-}16\times8\text{-}1}$ & $175$ & $274925$ & $0$ & $0$ & $0$ & $0$  & $138$ \\
${784\text{-}32\times2\text{-}1}$ & $192$ & $301632$ & $0$ & $0$ & $0$ & $15$ & $4436$ \\
${784\text{-}64\times2\text{-}1}$ & $192$ & $301632$ & $0$ & $0$ & $0$ & $0$  & $10385$ \\
\bottomrule
\end{tabular}
\end{table}

\Cref{fig:weightdifferences} compares the distributions of coordinate-wise weight differences between pairs of first-layer neurons in models trained with and without \nospacetd.
The distributions for models trained with \td show a greater concentration of small differences, indicating that the similarity penalty brings the weight vectors closer together.
We then examine whether this increased weight similarity leads to joint activation changes under small perturbations around the critical points found by the attack.
We examine whether perturbations around the critical points returned by the attack cause multiple neurons to change activation states.
As shown in \Cref{tab:joint-activation}, no probes cross multiple first layer boundaries at step sizes $10^{-5.5}$, $10^{-5}$, or $10^{-4.5}$ across the six protected models.
At $\varepsilon=10^{-4}$, only the $32\times2$ model exhibits such crossings, with $15$ occurrences among $301{,}632$ probes.
The remaining five models exhibit no multiple crossings at this step size.
Increasing the step size to $10^{-3}$ produces multiple crossings in all six models, with counts ranging from $19$ to $10{,}385$.
Since the same critical points and perturbation directions are used across step sizes, this increase reflects the effect of wider probes crossing additional boundaries.
The results show that the weight similarity induced by \td does not lead to frequent joint activation changes at the inspected critical points under small perturbations.

\subsection{RQ2: Clustering Errors in the Original Attack}\label{sec:experiments:rq2}
\begin{table}[t]
\centering
\caption{Clustering results of the original attack on normal and \td models. \textbf{Size}, \textbf{IDs}, and \textbf{Edges} denote the number of neurons incorrectly clustered together, their indices, and the resulting erroneous clustering edges.}
\label{tab:original_clustering}
\setlength{\tabcolsep}{3pt}
\renewcommand{\arraystretch}{1.05}
\begin{tabular}{clcccr}
\toprule
\multirow{2}{*}{\textbf{Model Mode}}
& \multirow{2}{*}{\textbf{Model}}
& \multirow{2}{*}{\textbf{Width}}
& \multicolumn{3}{c}{\textbf{Error Clustering}} \\
\cmidrule(lr){4-6}
& & & \textbf{Size} & \textbf{IDs} & \textbf{Edges} \\
\midrule
\multirow{4}{*}{Normal}
& ${784\text{-}8\times2\text{-}1}$
& $8$& $0$& N/A& $0$ \\
&${784\text{-}16\times2\text{-}1}$& $16$& $0$& N/A& $0$ \\
& ${784\text{-}32\times2\text{-}1}$& $32$& $0$& N/A& $0$ \\& ${784\text{-}64\times2\text{-}1}$& $64$& $0$& N/A& $0$ \\
\midrule
\multirow{4}{*}{\td}& ${784\text{-}8\times2\text{-}1}$& $8$& $8$& $0,1\ldots 6,7$& $83$ \\
&${784\text{-}16\times2\text{-}1}$& $16$& $16$& $0,1\ldots 14,15$& $337$ \\
& ${784\text{-}32\times2\text{-}1}$& $32$& $32$& $0,1\ldots 30,31$& $12338$ \\
&${784\text{-}64\times2\text{-}1}$& $64$& $64$& $0,1\ldots62,63$& $50074$ \\
\bottomrule
\end{tabular}%
\end{table}

We examine how \td affects the assignment of local observations to individual neurons.
\Cref{tab:original_clustering} reports the clustering results for ordinary and protected models with hidden widths $8$, $16$, $32$, and $64$.
No erroneous connections are reported for the ordinary models.
For the protected models, incorrect clustering involves all $8$, $16$, $32$, and $64$ neuron identities.
The corresponding numbers of erroneous edges are $83$, $337$, $12{,}338$, and $50{,}074$.
These connections show that the original matching rule accepts observations from distinct neurons as belonging to the same neuron.
Because the attack forms clusters through graph connectivity, such connections can also merge otherwise separate groups of observations.
The results support the mechanism analyzed in \Cref{sec:clustering-failure}: agreement on sufficiently many normalized coordinates does not reliably establish neuron identity when weights are similar.
Signature clustering is therefore a concrete failure point in the evaluated attack, motivating the use of a criterion that checks consistency across the observed weights and biases.

\subsection{RQ3: Signature Recovery with \textbf{\textsf{SimBreak}}}\label{sec:experiments:rq3}

We evaluate whether \ours enables signature recovery in the protected models.
As shown in \Cref{tab:ada_attack}, the original attack fails to cluster correctly and reports a recovery rate of $0\%$ across all six models.
With \ours, clustering succeeds, and the recovery rate reaches $100\%$ for the first layer in every configuration.
The relative errors range from $2^{-24.04}$ to $2^{-22.27}$.
For networks with two hidden layers, complete recovery is maintained as the width increases from $8$ to $64$.
The $16\times3$ and $16\times8$ models also achieve complete first layer signature recovery, with relative errors of $2^{-23.83}$ and $2^{-23.98}$, respectively.
These results demonstrate a bypass of the defense at the protected layer signature recovery stage across every evaluated configuration.
By checking direction and bias consistency, \ours~separates local signature observations that the original criterion incorrectly associates, turning the original attack's $0\%$ recovery into $100\%$ recovery without modifying the defended models.

\begin{table}[t]
\centering
\caption{First layer signature recovery on FCNs protected by $\mathsf{TraintoDefend}.$ \ours achieves complete signature recovery across all six models. N/A indicates unavailable measurements for unsuccessful runs.}
\setlength{\tabcolsep}{4pt}
\label{tab:ada_attack}
\begin{tabular}{ccccccccc}
\toprule
\makecell[c]{\bfseries Clustering\\ \bfseries Rule}
&\bfseries Structure
&\makecell[c]{\bfseries Correct\\ \bfseries Clustering}
&\makecell[c]{\bfseries Recovery\\ \bfseries Rate (\%)}
&\makecell[c]{\bfseries Relative\\ \bfseries Error}
&\makecell[c]{\bfseries Critical\\ \bfseries Points}
\\
\midrule
\multirow{6}{*}{\shortstack{Coordinate \\Agreement}}
&$784\text{-}8\times2\text{-}1$&\no&$0.00$&N/A&N/A\\
&$784\text{-}16\times2\text{-}1$&\no&$0.00$&N/A&N/A\\
&$784\text{-}32\times2\text{-}1$&\no&$0.00$&N/A&N/A\\
&$784\text{-}64\times2\text{-}1$&\no&$0.00$&N/A&N/A\\
&$784\text{-}16\times3\text{-}1$&\no&$0.00$&N/A&N/A\\
&$784\text{-}16\times8\text{-}1$&\no&$0.00$&N/A&N/A\\
\midrule
% \multirow{6}{*}{Adaptive}
\multirow{6}{*}{\shortstack{\ours \\\bfseries (Ours)}}
&$784\text{-}8\times2\text{-}1$&\yes&$100.00$&$2^{-24.04}$&$2^{4.75}$\\
&$784\text{-}16\times2\text{-}1$&\yes&$100.00$&$2^{-23.81}$&$2^{5.70}$\\
&$784\text{-}32\times2\text{-}1$&\yes&$100.00$&$2^{-23.66}$&$2^{6.60}$\\
&$784\text{-}64\times2\text{-}1$&\yes&$100.00$&$2^{-22.27}$&$2^{7.64}$\\
&$784\text{-}16\times3\text{-}1$&\yes&$100.00$&$2^{-23.83}$&$2^{5.64}$\\
&$784\text{-}16\times8\text{-}1$&\yes&$100.00$&$2^{-23.98}$&$2^{5.81}$\\
\bottomrule
\end{tabular}
\end{table}

\section{Discussion}
\label{sec:discussion}
Our attack challenges the protection offered by the first dedicated defense against cryptanalytic parameter extraction.
Across the evaluated models, the defense's reported barrier to protected layer signature recovery disappears once the attacker adapts the clustering criterion.
A defense may disrupt assumptions embedded in an existing implementation while leaving useful parameter information accessible through model outputs.
In the evaluated models, replacing the original clustering criterion enables protected layer signature recovery without changing the defended networks.
The failure of an unchanged attack therefore does not establish resistance to adaptive extraction.
Defenses based on neuron similarity must account for attacks that exploit both weight directions and affine offsets, rather than relying on the coordinate matching assumptions of an existing implementation.

\section{Conclusion}
\label{sec:conclusion}
In this paper, we challenge \nospacetd's premise that weight similarity prevents cryptanalytic extraction by causing neurons to switch activation states together.
Joint activation changes are rare under small perturbations around the critical points, while our analysis identifies a clustering obstacle caused by erroneous matches between distinct neuron signatures.
To overcome this obstacle, we propose \ours{}, which jointly assesses weight direction and bias consistency under a relative scale fitted over all observed weight coordinates.
Across six models, \ours{} raises signature recovery from $0\%$ to $100\%$ in the protected layer, with maximum relative error of $2^{-22.27}$.
These results demonstrate that \ours{} can bypass the signature recovery barrier imposed by \td and underscore the need to evaluate defenses against adaptive attacks.

\bibliographystyle{splncs04}
\bibliography{main.bib}
\appendix

\end{document}